\documentclass[prd,aps,nofootinbib
]{revtex4}
\usepackage{graphicx,epsf,amsmath,amsfonts,amssymb,amsbsy}
\newcommand{\vev}[1]{\langle#1\rangle}
\newcommand{\mat}{\left ( \begin{array}}
\newcommand{\emat}{\end{array} \right )}
\newcommand{\vect}{\left ( \begin{array}{c}}
\newcommand{\evect}{\end{array} \right )}

\begin{document}

\title{ \bf Does there arise a significant enhancement 
of the dynamical quark mass in external magnetic field?}
\author{
K.G.~Klimenko$^{\dagger,\ddagger}$, V.Ch. Zhukovsky$^{\flat}$}
\affiliation{$^{\dagger}$ Institute
for High Energy Physics, 142281, Protvino, Moscow Region, Russia}
\affiliation{$^{\ddagger}$ Dubna University (Protvino
branch), 142281, Protvino, Moscow Region, Russia}
\affiliation{$^{\flat}$ Faculty of Physics, Moscow State University,
119991, Moscow, Russia}

\begin{abstract}
Recently, it was found in QED that the generation of a dynamical
electron mass in a strong magnetic field is significantly enhanced by
the perturbative electron mass. In the present paper, the related
question of a possible enhancement of the dynamical quark mass in an
external magnetic field and with a bare mass term is investigated in
the Nambu--Jona-Lasinio model.
\end{abstract}

%\pacs{11.30.Qc, 12.39.-x, 21.65.+f}
% 11.30.Qc Spontaneous and radiative symmetry breaking
% 12.39.-x Phenomenological quark models
% 11.15.Ex Spontaneous breaking of gauge symmetries
% 12.38.Aw General properties of QCD
% 12.38.Mh Quark-gluon plasma
% 11.10.St Bound and unstable states; Bethe-Salpeter equations
% 12.38.-t Quantum chromodynamics
% 12.38.Lg Other nonperturbative calculations
% 26.60.+c Nuclear matter aspects of neutron stars
% 21.65.+f Nuclear matter
% 12.39.Fe

%\keywords{Gross -- Neveu model; pion condensation;
%Nambu--Goldstone bosons}
\maketitle
%\draft
%\large
%\maketitle

%\section{Introduction}

In the recent paper \cite{wang} some aspects of the well-known
magnetic catalysis effect were studied in quantum electrodynamics
(QED). In particular, it was shown that an enormously high external
magnetic field $B_{m_e}$ of the order of $10^{82}$ G would be needed
to create dynamically the common experimental value $m_e\approx 0.5$
MeV of the electron mass $m_e$ in the massless QED.  At the same
time, if the bare mass of the theory is nonvanishing and corresponds
to the experimental value $m_e$, then in the presence of an external
magnetic field of the same value $B_{m_e}$  the dynamical
mass of an electron is enhanced to the value almost ten times larger
than $m_e$. 
(The behaviour of the dynamical electron mass in a strong magnetic
field in massive QED was also considered previously \cite{gus}.)
As it was claimed in \cite{wang}, such a
significant enhancement of the dynamical electron mass in QED is a
new effect that can find applications in astrophysics and
cosmology, and it deserves to be investigated in more detail, and
especially also in QCD. In particular, as it was predicted in
\cite{wang}, in the case with much smaller and realistic magnetic
field values around $10^{15}$ G, the typical magnetic fields of
compact stars, a few percent measurable increase in the dynamical
electron mass still exists.

It has been known that the magnetic catalysis effect, i.e. the
spontaneous breaking of the chiral symmetry induced by an external
magnetic field $B$, is an universal phenomenon, which takes place in
different physical models (see, e.g., the reviews \cite{mir} as
well as the original papers
\cite{klim,gusynin,mikheev,gus2,wang2,bashir,sem} and
references therein). Thus, the following natural question arises:
{\bf Does it mean that the enhancement effect has an universal
character as well?} 
In this paper, we study this problem in the framework of the Nambu --
Jona-Lasinio (NJL) model with two quark flavors. 

In four-dimensional spacetime and at $B=0$ the system is
described by the following Lagrangian:
\begin{eqnarray}
 {\mathcal L}=\bar q[\mbox{i}\gamma^\nu \partial_\nu-m_0]q
  +G[(\bar qq)^2+(\bar qi\gamma^5\vec \tau q)^2],
  \label{1}
\end{eqnarray}
where the quark field $q\equiv q_{i\alpha}$ is a flavor doublet
($i=1,2$ or $i=u,d$) and a color triplet ($\alpha=1,2,3$ or
$\alpha=r,g,b$) as well as a four-component Dirac spinor; 
$\tau_a$ stands for the Pauli matrices. It is supposed here that up
and
down quarks have an equal current (bare) mass $m_0$. Clearly, at
$m_0=0$ this Lagrangian is invariant under the continuous chiral
SU(2)$_L\times$SU(2)$_R$ group as well as under the discrete chiral
transformation, $q\to i\gamma^5 q$. At the tree level, the
Lagrangian (\ref{1}) contains two free model parameters, the
coupling constant $G$ and the bare quark mass $m_0$. However,
when including quantum effects (quark loops), one should regularize
the corresponding loop integrals, for example, by cutting off the
three-dimensional momentum space, i.e. supposing that $|\vec
p|\le\Lambda$. Thus an additional free parameter, the cutoff
$\Lambda$, appeares in the model.
In the mean field approximation the effective potential of the model
(\ref{1}) looks like (see, e.g., \cite{buballa,ebert})
\begin{eqnarray}
&&V(m)=\frac{(m-m_0)^2}{4G}
-\frac{3}{4\pi^2}\left
[\Lambda(2\Lambda^2+m^2)\sqrt{m^2+\Lambda^2}
-m^4\ln\left (\frac {\Lambda+\sqrt{m^2+\Lambda^2}}{m}\right
)\right ],
\label{2}
\end{eqnarray}
where $m$ is the dynamical quark mass, which is connected with the
bare mass $m_0$ and the vacuum expectation value of quark fields
$\vev{\bar qq}$ through the relation 
\[m=m_0-2G\vev{\bar qq}.\]
Note that it depends on the model parameters $G,m_0,\Lambda$ and is
determined by the gap equation 
\begin{equation}
\frac\partial{\partial m}
V(m)\equiv\frac{m-m_0}{2G}-\frac{3m}{\pi^2}\left
[\Lambda\sqrt{m^2+\Lambda^2}
-m^2\ln\left (\frac {\Lambda+\sqrt{m^2+\Lambda^2}}{m}\right
)\right ]=0. \label{3}
\end{equation}
Evidently \cite{buballa,ebert}, at $m_0=0$ the dynamical quark mass
$m$ is a nonzero quantity only at $G>G_{crit}=\pi^2/(6\Lambda^2)$
(in this case the chiral symmetry of the model is spontaneously
broken down). However, it follows from (\ref{3}) that at $m_0=0$ and
$G<G_{crit}$ we have $m\equiv 0$, and the chiral symmetry remains
intact in this case. If $m_0\ne 0$, then $m\ne 0$ for arbitrary
values of $G$. Below, one can find some values of $m$ vs $m_0$ in the
second line of Tables I, II for $G<G_{crit}$. 

The influence of an external constant and homogeneous magnetic field
$B$ on the properties of the NJL-type models was already considered
in refs. \cite{ebert,shovkovy,njl,inagaki}. To obtain the 
corresponding Lagrangian, it is necessary to perform in (\ref{1}) the
following replacement: $\partial_\nu\to\partial_\nu+\mbox{i}QA_\nu$,
where $A_\nu$ is a vector-potential of an external magnetic field
$B$, and $Q=\mbox{diag} (e_1,e_2)$ is the electric charge matrix of
quarks. Here $e_1=2|e|/3$ and $e_2=-|e|/3$ ($e$ is the electric
charge of electrons) are the electric charges of $u$- and $d$-
quarks, respectively. At $m_0=0$, the resulting Lagrangian is still
symmetric with respect to the discrete symmetry $q\to i\gamma^5 q$,
but it is no more invariant under the continuous chiral symmetry
SU(2)$_L\times$SU(2)$_R$ because of the difference in quark
electric charges. Clearly, at $B\ne 0$ the effective potential of the
model is also modified (see, e.g., \cite{ebert}) and looks like
\begin{equation}
\label{4}
V_{\rm eff}(M;B)=V(M)-
\sum_{i=1}^2\frac{3(e_iH)^2}{2\pi^2}\Bigl\{\zeta
'(-1,x_i)-\frac 12[x_i^2-x_i]\ln x_i
+\frac{x_i^2}4\Bigr\},
\end{equation}
where $x_i=M^2/(2|e_iB|)$ for each $i=1,2$,
$\zeta'(-1,x)$$=d\zeta(\nu,x)
/d\nu|_{\nu=-1}$ ($\zeta (\nu,x)$ is the generalized Riemann
zeta-function \cite{bateman}),  and $V(M)$ is the
effective potential (\ref{2}) with $m$ replaced by $M$.
The quantity $M=M(m_0,B)$ in (\ref{4}) is the
dynamical quark mass which is the solution of the gap equation
\begin{equation}
\label{5}
\frac{\partial}{\partial M}V_{\rm eff}(M;B)
\equiv\frac{\partial}{\partial M}V(M)-I_1(M)-I_2(M)=0,
\end{equation}
where 
\begin{eqnarray}
\label{6}
I_i(M)=\frac{3M|e_iB|}{2\pi^2}\{\ln\Gamma(x_i)-\frac
12\ln(2\pi)+x_i-\frac 12(2x_i-1)\ln x_i\} \,\,\,\,(i=1,2)
\end{eqnarray}
and $\Gamma(x)$ is the Euler gamma-function
\cite{bateman}. 
In what follows, we suppose that $B$ is a nonnegative quantity,
and take into consideration that $M(m_0,0)=m$.

At zero bare mass $m_0=0$ the influence of an external magnetic field
on the phase structure of the model (\ref{1}) was already
investigated, e.g., in \cite{ebert,shovkovy,njl,inagaki}. In
particular, it was shown there that at $G<G_{crit}$ and $B=0$ the
global minimum of the effective potential (\ref{2}) lies at the point
$m=0$, so that the chiral symmetries, both continuous and discrete,
are not broken down. However, at arbitrary small values of $B$ and
$G<G_{crit}$ the global minimum of the effective potential (\ref{4})
of the system is shifted to a nontrivial point. As a result, in this
case the spontaneous breaking of the discrete chiral symmetry 
\footnote{The continuous chiral symmetry remains to be 
broken due to the presence of the isospin-violating electric charge
matrix $Q$ in the covariant derivative of the modified Lagrangian
(see the remarks above (\ref{4})).}
is induced by an external magnetic field $B\ne 0$ (magnetic catalysis
effect). Moreover, a dynamical quark mass $M\equiv M(m_0=0,B)$,
which is the solution of the equation (\ref{5}) at $m_0=0$, is also
generated. Note, at $G>G_{crit}$ and $m_0=0$ the dynamical chiral
symmetry breaking in the NJL model takes place even at $B=0$ due to a
rather strong interaction in the quark-antiquark channel.

Now, we have at our disposal all necessary formulas in order to solve
the question raised at the beginning of the paper. 
Clearly, the two different possibilities should be studied, 
$G<G_{crit}$ and $G>G_{crit}$.

\underline{The weak coupling regime ($G<G_{crit}$).} Since in this
case, as in QED, the magnetic catalysis effect takes place in the NJL
model, i.e. at $m_0=0$ a nonzero dynamical quark mass is induced by
an external magnetic field, we are going to proceed in the
spirit of the paper \cite{wang}. For simplicity, let us put
$\Lambda=1$ GeV, i.e. $G_{crit}\approx 1.65$ GeV$^{-2}$, and
consider, for illustrations, two values of the coupling constant,
$G=0.5$ GeV$^{-2}$ and $G=1$ GeV$^{-2}$, which are smaller, than
$G_{crit}$. Now, in order to compare our results with those of Wang,
it is convenient to divide, as in \cite{wang}, the numerical
calculations into several stages.

\vspace{0.5cm}

\begin{minipage}{0.49\hsize} 
\begin{center}
Table I: The case $G=0.5$ GeV$^{-2}$.
\end{center}

\begin{tabular}{|c|c|c|c|c|c|} \hline
 $m_0$~[GeV] &
  0 & 0.0003 & 0.003 &
 0.03 & 0.3   \\ \hline
$m$~[GeV] &
0 & 0.00043 & 0.0043 &
0.043 & 0.4  \\ \hline
$2|e|B_{m}$~[GeV$^2$] & 0 & 1.38 & 2.01 & 3.59 & 10.24 \\
\hline
$M(m_0,B_{m})$~[GeV] &
0 & 0.00194 & 0.0154 &
0.113 & 0.74  \\ \hline
$R=M(m_0,B_{m})/m$ &
 & 4.51 & 3.58 & 2.64 & 1.85  \\ \hline
%\label{tab}
\end{tabular}
%\hfill
%\end{center}
%\end{table}
\end{minipage}\hfill
\begin{minipage}{0.49\hsize}
\begin{center}

Table II: The case $G=1$ GeV$^{-2}$.
\end{center}

\begin{tabular}{|c|c|c|c|c|c|} \hline
 $m_0$~[GeV] &
  0 & 0.0002 & 0.002 &
 0.02 & 0.2   \\ \hline
$m$~[GeV] &
0 & 0.0005 & 0.0051 &
0.051 & 0.4  \\ \hline
$2|e|B_{m}$~[GeV$^2$] & 0 & 0.435 & 0.67 & 1.36 & 4.86 \\
\hline
$M(m_0,B_{m})$~[GeV] &
0 & 0.00212 & 0.0165 &
0.116 & 0.65  \\ \hline
$R=M(m_0,B_{m})/m$ &
 & 4.16 & 3.24 & 2.30 & 1.64  \\ \hline
\end{tabular}
%\hfill
\end{minipage}
\vspace{0.5cm}

(i) First, we put $B=0$ and find the dynamical quark mass
$m$, by solving  equation (\ref{3}) for different values of 
the bare mass $m_0$. For some representative values of $m_0$ (see the
first line in Tables I, II) the corresponding values of $m$ are
presented in Tables I, II (see the second line there). 

(ii) Next, one should find such a value of the magnetic field
$B_{m}$, for which the solution of equation (\ref{5}) at $m_0=0$ 
coincides with $m$, i.e. $M(m_0=0,B_{m})=m$. For
the values of $m$ from Tables I, II the corresponding values of
$B_{m}$ are presented in the third line of Tables I, II. Since
%%% More accurate number using: 1 GeV^2 =5.33 x 10^19 G !! %%%%%%%
%$\Lambda^2/|e|\approx 3$ GeV$^{2}\sim 3 10^{19}$ G,
$\Lambda^2/|e|\approx 3$ GeV$^{2}\approx 1.6\cdot 10^{20}$ G,
we see that in the NJL model the values of the quantity $B_{m}$ 
have the order of magnitude of the NJL characteristic magnetic scale
$B_c$, $B_c\equiv\Lambda^2/|e|$. In contrast, in QED (see
\cite{wang}) the values of the corresponding quantity $B_{m_e}$ are
unrealistically higher than the characteristic QED magnetic scale,
the Schwinger magnetic field $m^2_{e}/|e|\approx 4.4\cdot 10^{13}$ G.

(iii) Finally, for each fixed value of $m_0$ and $B_{m}$ we have
solved
the gap equation (\ref{5}) and found the corresponding dynamical
quark mass $M(m_0,B_{m})$ as well as the ratio $R\equiv
M(m_0,B_{m})/m$ (these quantities are given in the fourth and fifth
lines of Tables I, II respectively) which in some sense might serve
as a measure of the dynamical quark mass enhancement effect
\cite{wang}. It is seen from 
these tables that for the considered values of $m_0$ we have obtained
$R<5$ in the framework of the NJL model (\ref{1}), and even $R<2$ for
the physically interesting case of a dynamical quark mass $m=0.4$ GeV
(with bare quark mass $m_0=0.3$ GeV for $G=0.5$ GeV$^{-2}$ or
$m_0=0.2$ GeV for $G=1$ GeV$^{-2}$). For comparision, let us quote
the value $R\approx 10$, which was obtained in the same manner in QED
\cite{wang}. 

Judging about the possibility of the enhancement effect in terms
of the quantity $R$ only, one might conclude that in the framework of
the NJL model with a rather weak interaction, $G<G_{crit}$, the
generation of a dynamical quark mass in a strong magnetic field 
is still enhanced at nonzero bare quark mass (since $R\approx 2$ in
physically reliable cases with $m\approx 0.4$ GeV). But here this
effect is not so pronounced as in QED, where $R\approx 10$.
On the other hand, one should keep in mind that in the NJL model the
enhancement of the dynamical quark mass takes place only at
sufficiently high magnetic fields $B\gtrsim B_m\sim 10^{20}$ G.
Indeed, our calculations show that in a more interesting case with 
realistic values of an external magnetic field $B\lesssim
B_{phys}\equiv 10^{15}$ G, which are typical values of magnetic
fields on the surface of young neutron stars, the dynamical quark
mass $M(m_0\ne 0,B)$ is with great accuracy equal to the dynamical
(constituent) quark mass $m=M(m_0\ne 0,0)$ at $B=0$. As a result, we
see that for realistic values of $B$ the enhancement effect is
absent. In contrast, in QED such magnetic fields provide a few
percent increase of the dynamical electron mass in comparision with
$m_e$ at $B=0$, which is sufficient for observation of this effect in
experiments \cite{wang}. To better understand the absence of the
enhancement effect in the NJL model at $B\lesssim B_{phys}$, one
should take into account that in the framework of the NJL model the
field $B_{phys}$ is comparatively weak, since $B_{phys}\ll B_c$,
while in QED the field $B_{phys}$ is comparatively strong, since
$B_{phys}\gg m_e^2/|e|$, i.e. it is much greater than the 
Schwinger field.

\underline{The strong coupling regime ($G>G_{crit}$).}
Since NJL models are
considered to be effective theories for low energy QCD only at
$G>G_{crit}$, we have studied the influence of an external magnetic
field $B$ on the dynamical quark mass also in this case. At
$G>G_{crit}$ and $B=0$, the values of the NJL model parameters can be
fixed through fitting of experimental data, and the typical set of
$\Lambda,m_0,G$ looks like \cite{buballa}: $\Lambda=0.6$ GeV,
$m_0=0.005$ GeV, $G=6.73$ GeV$^{-2}$, which corresponds to
$G_{crit}\approx 4.57$ GeV$^{-2}$ and the dynamical quark mass
$M(m_0,B=0)\approx 0.4$ GeV. 

Now, using the gap equation (\ref{5}), it is possible to conclude
that at the value of the external magnetic field $B=B_c\equiv
\Lambda^2/|e|\approx 6.4\cdot 10^{19}$ G, which is the characteristic
magnetic scale of the model for the above chosen value of
$\Lambda=0.6$ GeV, the dynamical quark mass $M(m_0,B_c)$ exceeds the
dynamical quark mass $M(m_0,B=0)$ no more than by 20$\%$. At
$B\approx 25B_c$, the corresponding dynamical quark mass is ten times
larger than $M(m_0,B=0)$, etc. Hence, at $G>G_{crit}$, and in a
rather strong external magnetic field $B\gtrsim B_c$ the enhancement
of the dynamical quark mass also takes place.

However, for values of a magnetic field $B$ smaller than $B_c$, the
excess of $M(m_0,B)$ over $M(m_0,B=0)$ sharply decreases.
(Note, that it is just in the region $B\lesssim B_c$ that the
dynamics of QCD is qualitatively similar to that in the NJL model
\cite{shovk}.) 
Indeed, at $B=0.1B_c$ it is equal to 0.3$\%$ etc., and for the value 
$B=B_{phys}\equiv 10^{15}$ G (the field on the surface of young
neutron stars) the difference between $M(m_0,B_{phys})$ and
$M(m_0,B=0)$ starts from the 10-th significant digit, i.e. it is
vanishingly small. Therefore, for sufficiently small $B\lesssim
B_{phys}\ll B_c$ the enhancement effect is absent both in the NJL
model and QCD 
\footnote{The behaviour of the quark condensate $\Sigma(B)$ at small
values of magnetic fields $B\ll B_c$ was also considered in the
framework of the chiral effective theory \cite{cohen}. It is easily
seen that in this case at $B\lesssim B_{phys}$ the quark condensate
$\Sigma(B)$ exceeds $\Sigma(0)$ also in sufficiently small fractions
of a percent and, hence, the dependence of chiral condensate on $B$
might not be allowed for.}, and hence, in physical applications one
might ignore the dependence of the dynamical quark mass on 
an external magnetic field $B$ in this range. In spite of this fact,
there are other phenomena, which can be observed at $B\lesssim
B_{phys}$ in dense quark matter. Among them are the magnetic
oscillation effect and other effects \cite{ebert,njl} that are not
directly related to the behaviour of the dynamical quark mass vs $B$.
They are connected mostly with the thermodynamical properties of the
system. 

We remark in conclusion, that in order to answer the question raised
at the beginning of the paper, one should first establish the ranges
for the external magnetic field $B$ under consideration. Then, if $B$
varies in a certain vicinity of $B_{phys}\equiv 10^{15}$ G, the
considered enhancement effect is intrinsic to QED (since here the
magnetic field $B_{phys}$ can be considered to be strong enough). At
the same time, in  QCD  or the NJL model, the dynamical quark
mass is not influenced by these realistic values of an external
magnetic field. However, if $B$ is rather strong, i.e., $B\gtrsim
B_c$, the enhancement of a dynamical fermion mass does occur in 
QED and the NJL-type models, etc. 

Note that in QCD in a strong magnetic field the situation with the
enhancement effect might be very involved. Indeed, as it was shown in
\cite{shovk} in the chiral limit of QCD, the dynamical quark mass at
$B\gtrsim B_c$ behaves quite unexpectedly since in a wide range of
strong magnetic fields it is suppressed in comparision with the
dynamical quark mass at $B=0$ (one of the reasons is that QCD is an
asymptotically free model). The same behaviour of the dynamical quark
mass might be inherent to QCD with nonzero bare quark mass. 
%So, in QCD in a strong magnetic field a special consideration of the
%enhancement effect is needed.

\section*{Acknowledgement}

The authors are grateful to D. Ebert for careful reading of
the manuscript and fruitful discussions as well as to V.P. Gusynin
and S.Y. Wang for useful correspondence and remarks.

\end{document}